\documentclass[a4paper,11pt]{article}

\usepackage{citesort}
\usepackage{epsfig}

\makeatletter

\@addtoreset{equation}{section}
\makeatother
\def\lsim{\;\raise0.3ex\hbox{$<$\kern-0.75em\raise-1.1ex\hbox{$\sim$}}\;}
\def\gsim{\;\raise0.3ex\hbox{$>$\kern-0.75em\raise-1.1ex\hbox{$\sim$}}\;}

\def\ben{\begin{enumerate}}  \def\een{\end{enumerate}}
\def\bit{\begin{itemize}}    \def\eit{\end{itemize}}
\def\beq{\begin{equation}}   \def\eeq{\end{equation}}
\def\ba{\begin{array}}       \def\ea{\end{array}}
\def\bea{\begin{eqnarray}}   \def\eea{\end{eqnarray}}
\def\nn{\nonumber}

\def\stq{\sin^2{\theta_W}}
\def\stqq{\sin^4{\theta_W}}
\def\sb{\sin{\beta}}
\def\sbq{\sin^2{\beta}}
\def\cb{\cos{\beta}}
\def\cbq{\cos^2{\beta}}
\def\sbb{\sin{2\beta}}

\def\sbbq{\sin^2{2\beta}}
\def\sbbqq{\sin^4{2\beta}}

\def\cbbq{\cos^2{2\beta}}

\def\LM{\mathrm{L_M}}
\def\Lmu{\mathrm{L}_\mu}
\def\Lnu{\mathrm{L}_\nu}
\def\LL1{\mathrm{L}_1}
\def\L2{\mathrm{L}_2}
\def\La{\mathrm{L}_\mathrm{A}}
\def\Ls{\mathrm{L}_\mathrm{S}}

\def\Lep{\mathrm{L}_{\mathrm{E}_+}}
\def\Lmas{\mathrm{L}_\mathrm{MAS}}
\def\Lepm{\mathrm{L}_{\mathrm{E}_+\mathrm{E}_-}}

\begin{document}

\begin{titlepage}
\renewcommand{\thefootnote}{\fnsymbol{footnote}}
\setcounter{footnote}{0}

\begin{flushright}
LPT Orsay 05-28 \\
\end{flushright}

\begin{center}
\vspace{3cm}
{\Large\bf Yukawa Induced Radiative Corrections to the \\ Lightest Higgs Boson
Mass in the NMSSM} \\
\vspace{2cm}
{\bf Ulrich Ellwanger\footnote{email: ellwanger@th.u-psud.fr} and Cyril
Hugonie\footnote{email: hugonie@th.u-psud.fr}} \\
Laboratoire de Physique Th\'eorique\footnote{Unit\'e mixte de Recherche -- CNRS
-- UMR 8627} \\
Universit\'e de Paris XI, F-91405 Orsay Cedex, France
\vspace{2cm}
\end{center}

\begin{abstract}
We compute the leading logarithmic radiative corrections to the lightest Higgs
mass in the NMSSM involving the electroweak gauge couplings and the NMSSM
specific Yukawa couplings $\lambda$ and $\kappa$ (including all mixed
combinations), which are induced by chargino, neutralino and Higgs boson loops.
The effect of the NMSSM specific Yukawa couplings $\lambda$ and $\kappa$ is to
increase the upper bound on the lightest Higgs mass by up to 
$\sim 1$~GeV, but they can also decrease the lightest Higgs mass by up
to $\sim -20$~GeV.
\end{abstract}

\end{titlepage}

\renewcommand{\thefootnote}{\arabic{footnote}}
\setcounter{footnote}{0}

\section{Introduction} \label{sec:int}

The perhaps most easily falsifiable prediction of supersymmetric extensions of
the Standard Model is the prediction of a relatively light Higgs boson. In the
MSSM, at tree level, the mass of the lightest CP even neutral Higgs boson is
bounded from above by $M_Z$, which would already be ruled out by negative Higgs
search results at LEP.

However, radiative corrections -- mainly due to loops of quarks and squarks of
the third generation -- lift this upper bound considerably. In the last years,
these radiative corrections to the mass of the lightest Higgs boson in the MSSM
have been computed to a fairly high precision (see refs.~\cite{r1,r2,r3} for
recent reviews).

The knowledge of the radiative corrections to the mass of the lightest Higgs
boson is important for the following reasons: Given either a lower experimental
bound on its mass, or a future measurement of its mass, they allow either to
decide whether a particular supersymmetric extension of the Standard Model is
ruled out, or to put constraints on additional parameters of the model (as the
squark masses or  $\tan\!\beta$) on which the radiative corrections depend.
(This resembles to the present indirect determination of the preferred Higgs
mass through precision tests in the standard model.) 

In the NMSSM, the upper bound on the mass of the lightest Higgs boson is
alleviated already at tree level. Subsequently we confine ourselves to the NMSSM
with a scale invariant superpotential~\cite{nmssm}
\beq
W = \lambda \widehat{S} \widehat{H}_u \widehat{H}_d + \frac{\kappa}{3} \,
\widehat{S}^3 + \ldots\ , 
\label{supot}
\eeq
which is the only supersymmetric extension of the Standard Model where the weak
scale originates from the soft susy breaking scale only, i.e. where no
supersymmetric dimensionful parameters as $\mu$ are present in the
superpotential. Now, the upper bound on the mass $M_h$ of the lightest CP even
Higgs is
\beq
M_h^2 \leq M_Z^2 \left ( \cos^2 \! 2\beta + \frac{2\lambda^2} {g_1^2+g_2^2} \,
\sin^2 \! 2\beta \right ) 
\label{tree}
\eeq
which depends on the new Yukawa coupling $\lambda$ (unless  $\sin^2 \! 2\beta$
is very small, i.e. $\tan\!\beta$ large). However, if one requires the absence
of a Landau singularity for $\lambda$ below the GUT scale, $\lambda$ is bounded
by $\sim 0.7$ from above~\cite{nmssm}, leading again to a tree level upper
bound  on the mass of the lightest Higgs boson that is, however, somewhat larger
than in the MSSM. Hence, as in the MSSM, the knowledge of the radiative
corrections to its mass are important.

At present, these radiative corrections in the NMSSM have not been computed to
the same accuracy as in the MSSM. Of course, radiative  corrections in the NMSSM
that are proportional to the quark/lepton Yukawa couplings and the gauge
couplings only are the same as in the MSSM, but there are many additional
contributions involving the new Yukawa couplings $\lambda$ and $\kappa$ in the
superpotential in eq.~(\ref{supot}), and the associated soft trilinear couplings
$A_\lambda$ and $A_\kappa$.

The one loop corrections in the NMSSM induced by t and b quark/squark loops have
been computed already some time ago~\cite{rad1}, and the dominant two loop
corrections ($\sim h_t^6$ and $\sim h_t^4 \alpha_s$), that are the same as in
the MSSM, have been included in an analysis of the NMSSM Higgs sector in
ref.~\cite{rad2}.

The leading logarithmic one loop corrections to the lightest Higgs mass in the
NMSSM proportional to the electroweak gauge couplings $g$ ($\sim g^4$, that are
the same as in the MSSM~\cite{habhemp}) are included in the code
NMHDECAY~\cite{nmhdecay}, where the NMSSM Higgs masses, couplings and branching
ratios are computed as functions of the parameters in the Lagrangian of the
model. Recall that large logarithmic one loop corrections $\sim g^4$ appear as
soon as the mass of a sparticle or a Higgs field is (much) larger than $M_Z$,
and they affect the mass of the lightest Higgs boson by $\sim -3$
GeV~\cite{habhemp}.

In this case, however, there appear also large logarithmic one loop corrections
in the NMSSM $\sim \lambda^4, g^2\lambda^2, \kappa^4$ etc, that are of the same
order as the pure electroweak corrections. It is the purpose of this paper to
compute these radiative corrections in the NMSSM defined by the superpotential
above.

The particles whose loops generate these contributions to the lightest
Higgs mass are the neutralinos, charginos and the CP even, CP odd and
charged Higgs fields. There are many fields and couplings in the NMSSM
that are not present in the MSSM, and which lead to the NMSSM specific
radiative corrections. Their contributions to the upper bound on the
lightest Higgs boson mass remain limited, however, to $\sim 1$~GeV, if
one requires the absence of a Landau singularity below
$M_{\mathrm{GUT}}$ for all Yukawa couplings $\lambda$, $\kappa$ and
$h_t$. On the other hand CP odd loop contributions to the effective
potential can decrease the lightest Higgs boson mass by up to $-20$~GeV.

In contrast to ref.~\cite{habhemp} we do not use a RG analysis in an effective
two Higgs doublet model below a scale $M_A$ (as possible in the MSSM), since
there are more possible mass scales in the NMSSM and the corresponding effective
low energy theory is considerably more involved. Instead, we compute explicitly
the contributions to the Coleman-Weinberg effective potential, the Higgs self
energies and the modified running of the electroweak gauge couplings induced by
heavy sparticles and/or Higgs fields. Wherever a comparison with the results of
ref.~\cite{habhemp} on the mass of the lightest CP even Higgs boson is possible,
our results agree.

In the next section we compute the contributions of neutralino and char\-gino
loops to all elements of the CP even Higgs boson mass matrix (a $3 \times 3$
matrix in the NMSSM). The corresponding contributions from Higgs loops are,
however, quite lengthy (and numerically not very important). Hence, here we just
give the contributions to the mass of the lightest Higgs doublet field; these
contributions can easily be translated back to contributions to the $2 \times 2$
Higgs doublet sector of the CP even Higgs boson mass matrix. For completeness we
add the contributions from the squark/slepton sectors to the modified running of
the electroweak gauge couplings, that contribute also to the mass of the light
Higgs doublet to ${\cal O}(g^4)$, but which are the same as in the
MSSM~\cite{habhemp}.

In section 3 we study the numerical impact of the new radiative
corrections on the mass of the lightest CP even Higgs boson by studying
the dependence of the individual fermionic and bosonic contributions as
functions of $\tan\!\beta$, and the total contributions as functions
of  $\lambda$ for some specific choices of the other parameters.

\section{Chargino/Neutralino/Higgs Loop Corrections in the NMSSM} 
\label{sec:radcors}

As stated in the introduction, we consider the NMSSM with a scale
invariant superpotential, which reads (for third generation quarks
only, and using the conventions of ref.~\cite{nmhdecay}):
\beq
W =
\lambda \widehat{S}\widehat{H}_u\widehat{H}_d + \frac{1}{3} \kappa
\widehat{S}^3 + h_t\widehat{Q}\widehat{H}_u\widehat{T}_R^c - h_b
\widehat{Q}\widehat{H}_d\widehat{B}_R^c \ .
\label{fullsupot}
\eeq 
Hereafter, hatted capital letters denote superfields, and unhatted capital
letters the corresponding (complex) scalar components. The corresponding soft
terms are
\bea
& -{\cal L}_\mathrm{soft} = m_\mathrm{H_u}^2 | H_u |^2 + m_\mathrm{H_d}^2 | H_d
|^2 + m_\mathrm{S}^2 | S |^2 +m_Q^2|Q^2| + m_T^2|T_R^2| + m_B^2|B_R^2| \nn \\
& + ( \lambda A_\lambda H_u H_d S + \frac{1}{3} \kappa A_\kappa S^3 +
h_t A_t Q H_u T_R^c - h_b A_b Q H_d B_R^c + \mathrm{h.c.}) \ .
\label{soft}
\eea
Subsequently we denote the neutral CP even Higgs vevs by $h_u$, $h_d$
and $s$. It is also convenient to introduce the quantities $\mu =
\lambda s$ and $\nu = \kappa s$.

First we consider the loop corrections induced by charginos and
neutralinos. These contribute to three distinct effects that have to be
added up:

i) First, they contribute to the Colemann-Weinberg effective potential (in the
$\overline{\mathrm{DR}}$ scheme which is, however, irrelevant for the leading
logarithms)
\beq
\Delta V_{\mathrm{eff}} = \frac{1}{64\pi^2} \, \mbox{STr} M^4 \left [ \ln \left
( \frac{M^2}{Q^2} \right ) - \frac{3}{2} \right ] \ . 
\label{cw}
\eeq
Subsequently we will choose $Q^2 = M_Z^2$, hence all couplings and parameters in
eqs.~(\ref{fullsupot}) and (\ref{soft}) are renormalized at the scale $M_Z^2$.

ii) Second, they contribute to the Higgs self energies, which we can compute at
vanishing momenta within the present leading logarithmic approximation.

iii) Third, they affect the running of the electroweak gauge couplings at scales
below their (susy breaking) masses: Only at scales above all susy breaking
masses we can identify the electroweak gauge couplings with the parameters
$\hat{g}_1$, $\hat{g}_2$ that appear as D-terms in the tree level Higgs
potential in the form $(h_u^2 - h_d^2)^2(\hat{g}_1^2+\hat{g}_2^2)/8$. Quantum
effects contributing to the Higgs potential (in terms of properly normalized
fields) are taken care of through the contributions i) and ii) above, but at the
end we want to express the Higgs masses in terms of electroweak gauge couplings
at the scale $M_Z$ (and not at $M_{\mathrm{susy}}$). This latter effect has to
be treated separately.

The mass matrices $M_{\chi^\pm}$ and $M_{\chi^0}$ for the charginos and
neutralinos are (using the conventions of ref.~\cite{nmhdecay})
\beq
M_{\chi^\pm} = \left(\ba{cc} M_2 & g_2 h_u \\ g_2 h_d & \mu \ea\right) 
\label{charg}
\eeq
and
\beq
M_{\chi^0} =
\left( \ba{ccccc}
M_1 & 0 & \frac{g_1 h_u}{\sqrt{2}} & -\frac{g_1 h_d}{\sqrt{2}} & 0 \\
& M_2 & -\frac{g_2 h_u}{\sqrt{2}} & \frac{g_2 h_d}{\sqrt{2}} & 0 \\
& & 0 & -\mu & -\lambda h_d \\
& & & 0 & -\lambda h_u \\
& & & & 2 \nu
\ea \right) 
\label{neutr}
\eeq
respectively. Here $M_1$ and $M_2$ are soft susy breaking $U(1)$ and $SU(2)$
gaugino masses.

Large logarithms appear as soon as any of the charginos and/or neutralinos has a
mass much larger than $M_Z$. Accordingly we can expect the presence of the
following potentially large logarithms (here we do not distinguish between $M_1$
and $M_2$, whose ratio we do not consider as exponentially large):
\bea
\LM &=& \ln(M_2^2/M_Z^2) \sim \ln(M_1^2/M_Z^2)\nn \\
\Lmu &=& \ln(\mu^2/M_Z^2)\nn \\
\Lnu &=& \ln(4\nu^2/M_Z^2)\nn \\
\LL1 &=& \ln(\mathrm{Max}(M_2^2,\mu^2)/M_Z^2)\nn \\
\L2 &=& \ln(\mathrm{Max}(\mu^2,4\nu^2)/M_Z^2)\ .
\label{logs}
\eea

Next we give the sum of the three contributions i), ii) and iii) above to the 6
matrix elements $M^2_{uu}$, $M^2_{dd}$, $M^2_{ud}$, $M^2_{ss}$,  $M^2_{us}$,
$M^2_{ds}$ of the neutral CP even Higgs mass matrix in the NMSSM (in an obvious
notation). To this end we define $g^2 = \frac{g_1^2+g_2^2}{2}$, $v^2 = h_u^2 +
h_d^2$ (hence $M_Z^2 = g^2v^2$) and $\tan\!\beta = h_u/h_d$. Then we obtain
\bea
\Delta M^2_{uu} &=& \frac{1}{16\pi^2}\left( -\frac{8}{3}g^4 v^2 \sbq
\cos^4\theta_W\ \LM\right. \nn \\
&&\left. +\left[-\frac{4}{3}g^4 v^2 \sbq\ (2\stqq -2\stq+1) +2 \lambda \kappa
\mu^2 \cot\beta \right] \Lmu\right. \nn \\
&&\left. +2\lambda \kappa \nu^2 \cot\beta\ \Lnu + \left[g^4 v^2 \sbq\ (-16\stqq
+28\stq-14)\right.\right. \nn \\
&&\left.\left. + g^2 \mu \nu \cot\beta \ (-2\stq + 3)\right]\ \LL1 - 3\lambda
\kappa \mu^2 \cot\beta\ \L2\right) \ ,\\
\Delta M^2_{dd} &=&\frac{1}{16\pi^2}\left( -\frac{8}{3}g^4 v^2 \cbq
\cos^4\theta_W\ \LM\right. \nn \\
&&\left. +\left[-\frac{4}{3}g^4 v^2 \cbq\ (2\stqq -2\stq+1) +2 \lambda \kappa
\mu^2 \tan\!\beta \right]\ \Lmu\right. \nn \\
&&\left. +2\lambda \kappa \nu^2 \tan\!\beta\ \Lnu + \left[g^4 v^2 \cbq\
(-16\stqq +28\stq-14)\right.\right. \nn \\
&&\left.\left. + g^2 \mu \nu \tan\!\beta \ (-2\stq + 3)\right]\ \LL1 - 3\lambda
\kappa \mu^2 \tan\!\beta\ \L2\right) \ ,\\
\Delta M^2_{ud} &=& \frac{1}{16\pi^2}\left( \frac{4}{3}g^4 v^2 \sbb\ 
\cos^4\theta_W\ \LM\right. \nn \\
&&\left. +\left[\frac{2}{3}g^4 v^2 \sbb\ (2\stqq -2\stq+1) -2 \lambda \kappa
\mu^2 \right]\ \Lmu\right. \nn \\
&&\left. -2\lambda \kappa \nu^2\ \Lnu + \left[g^4 v^2 \sbb\
(2\stq-5)\right.\right. \nn \\
&&\left.\left. +g^2 v^2 \lambda^2 \sbb\ (-2\stq +3) + g^2 \mu \nu  \ (2\stq -
3)\right]\ \LL1\right.\nn \\
&&\left.+\left[-g^2 v^2 \lambda^2 \sbb + 3\lambda \kappa \mu^2\right] \
\L2\right) \ ,\\
\Delta M^2_{ss} &=& \frac{1}{16\pi^2}\left( 8\mu^2\ (\kappa^2-\lambda^2)\ \Lmu
-16 \kappa^2\nu^2\ \Lnu\right)\ ,\\
\Delta M^2_{us} &=& \frac{v \mu}{16\pi^2}\left( 4 \lambda^2(\lambda\sb
-\kappa\cb)\ \Lmu +4 \kappa^2(\lambda\sb -\kappa\cb)\ \Lnu\right.\nn \\
&&\left. + g^2 (\lambda\sb + \kappa\cb)\ (4\stq-6)\ \LL1\right.\nn \\
&&\left.+ \lambda (-16\kappa^2\sb -2\lambda^2\sb +6\lambda \kappa \cb)\
\L2\right)\ ,\\
\Delta M^2_{ds} &=& \frac{v \mu}{16\pi^2}\left( 4 \lambda^2(\lambda\cb
-\kappa\sb)\ \Lmu +4 \kappa^2(\lambda\cb -\kappa\sb)\ \Lnu\right.\nn \\
&& \left.+ g^2 (\lambda\cb + \kappa\sb)\ (4\stq-6)\ \LL1\right.\nn \\
&&\left.+ \lambda (-16\kappa^2\cb -2\lambda^2\cb +6\lambda \kappa \sb)\
\L2\right)\ .
\eea

In the doublet sector of the neutral CP even Higgs sector one can identify a
state whose mass is bounded from above by $M_Z$ at tree level in the MSSM, and
by eq.~(\ref{tree}) in the NMSSM. The radiative corrections to this upper bound
are quite important, and are given by $\Delta M^2_{h} = \sbq \Delta M^2_{uu}
+\cbq \Delta M^2_{dd} + \sbb \Delta M^2_{ud}$. From the chargino and neutralino
loops in the NMSSM we obtain
\bea
&&\Delta^{\mathrm{ferm}} M_h^2 = \frac{v^2}{16\pi^2}\left(  -\frac{8}{3}g^4
\cbbq \cos^4\theta_W\ \LM\right. \nn \\ 
&&\left. -\frac{4}{3}g^4 \cbbq\ (2\stqq -2\stq+1)\  \Lmu\right. \nn \\
&&\left. + \left[g^4 \left\{ \sbbq\ (8\stqq -12\stq+2)
-16\stqq+28\stq-14\right\} \right.\right. \nn \\
&&\left.\left. + g^2 \lambda^2\sbbq\ (-2\stq + 3)\right] \LL1 - g^2
\lambda^2\sbbq\ \L2\right) \ .
\label{dmferm}
\eea
This and the previous results agree, for $\lambda \to 0$, with the ones in
ref.~\cite{habhemp}.

Now we turn to the contributions from the charged, neutral CP even and CP odd
Higgs loops. These contribute to the Coleman-Weinberg effective potential and to
the modification of the running of the electroweak gauge couplings, but their
contributions to the Higgs self energies do not give raise to large logarithms.
The corresponding tree level mass matrices $M^2_\pm$, $M^2_S$ and $M^2_P$ are,
respectively,
\beq
M^2_\pm =
\left(\ba{cc}
m_\mathrm{H_u}^2 + \mu^2 +\frac{g^2(h_u^2-h_d^2)}{2} +\frac{g_2^2h_d^2}{2}  &
h_u h_d\left(\frac{g_2^2}{2}-\lambda^2\right) +\mu (\nu +A_\lambda)\\
h_u h_d\left(\frac{g_2^2}{2}-\lambda^2\right) +\mu (\nu +A_\lambda)  &\ \
m_\mathrm{H_d}^2 + \mu^2 +\frac{g^2(h_d^2-h_u^2)}{2} +\frac{g_2^2h_u^2 }{2}
\ea\right) \ ,
\label{chargh}
\eeq
\bea
M^2_{S,11} &=& m_\mathrm{H_u}^2 + \mu^2 +\lambda^2 h_d^2
+\frac{g^2}{2}(3h_u^2-h_d^2)\nn \\
M^2_{S,22} &=& m_\mathrm{H_d}^2 + \mu^2 +\lambda^2 h_u^2
+\frac{g^2}{2}(3h_d^2-h_u^2)\nn \\
M^2_{S,12} &=& h_u h_d (2\lambda^2 - g^2) -\mu (A_\lambda +\nu)\nn \\
M^2_{S,33} &=& m_\mathrm{S}^2 +\lambda^2 (h_u^2+h_d^2) +6\nu^2-2\lambda\kappa
 h_u h_d +2\nu A_\kappa\nn \\
M^2_{S,13} &=& 2\lambda\mu h_u -\lambda h_d (2\nu + A_\lambda)\nn \\
M^2_{S,23} &=& 2\lambda\mu h_d -\lambda h_u (2\nu + A_\lambda)\ ,
\label{scalh}
\eea
\bea
M^2_{P,11} &=& m_\mathrm{H_u}^2 + \mu^2  +\frac{g^2}{2}(h_u^2-h_d^2)\nn \\
M^2_{P,22} &=& m_\mathrm{H_d}^2 + \mu^2  +\frac{g^2}{2}(h_d^2-h_u^2)\nn \\
M^2_{P,12} &=& \mu (A_\lambda +\nu)\nn \\
M^2_{P,33} &=& m_\mathrm{S}^2 +\lambda^2 (h_u^2+h_d^2) +2\nu^2+2\lambda\kappa
h_u h_d -2\nu A_\kappa\nn \\
M^2_{P,13} &=& \lambda h_d (A_\lambda-2\nu)\nn \\
M^2_{P,23} &=& \lambda h_u (A_\lambda-2\nu)\ .
\label{pscalh}
\eea

These mass matrices have to be used in the Coleman-Weinberg effective potential
eq.~(\ref{cw}) and in the computation of its derivatives w.r.t. $h_u$, $h_d$ and
$s$ that contribute to the neutral CP even mass matrix; only then they can be
simplified by applying the minimization equations of the tree level potential
\bea
m_\mathrm{H_u}^2 &=&  - \mu^2 -\lambda^2 h_d^2
+\frac{g^2}{2}(h_d^2-h_u^2)+\cot\beta \mu (A_\lambda + \nu)\ ,\nn \\
m_\mathrm{H_d}^2 &=&  - \mu^2 -\lambda^2 h_u^2
+\frac{g^2}{2}(h_u^2-h_d^2)+\tan\!\beta \mu (A_\lambda + \nu)\ ,\nn \\
m_\mathrm{S}^2 &=& -\lambda^2(h_u^2 + h_d^2) -2\nu^2 + 2\lambda\kappa h_u h_d
-\nu A_\kappa +\lambda A_\lambda \frac{h_u h_d}{s}
\label{softm}
\eea
in order to eliminate the soft masses squared. 

As before, large logarithms appear in the loop contributions to the neutral CP
even mass matrix, if any of the eigenstates of the three mass matrices above is
much larger than $M_Z$.

In the charged Higgs sector, the diagonalization of the $2 \times 2$ mass matrix
is straightforward. As in the MSSM, an eigenstate (apart from the Goldstone
boson) is much heavier than $M_Z$ if (using eq.~(\ref{softm}))
\beq
M_A^2 \equiv \mu (A_\lambda +\nu)\frac{(h_u^2+h_d^2)}{h_u h_d}
\label{ma}
\eeq
satisfies 
\beq
M_A \gg M_Z\ .
\label{malim}
\eeq 
(Here we introduced the MSSM-like notation $M_A$ for the mass of the heavy Higgs
doublet in the limit (\ref{malim}).) Accordingly a potentially large logarithm
is given by $\La = \ln(M_A^2/M_Z^2)$.

In the neutral CP even sector, the situation is considerably more involved,
since we have to deal with a $3 \times 3$ mass matrix. It is convenient to
perform first a preliminary approximate diagonalization by defining
$\widetilde{M}^2_S = {\cal O}^T M_S^2\ {\cal O}$ with
\beq
{\cal O} = \left(\ba{ccc} \cb & \sb & 0\\
-\sb & \cb & 0\\
0 & 0 & 1
\ea\right) \ .
\label{o}
\eeq
(After the computation of the derivatives of the effective potential
w.r.t. the vevs, the angle $\beta$ can be identified with
$\arctan(h_u/h_d)$ as in the MSSM.)

In the limit (\ref{malim}) we have $\widetilde{M}^2_{S,11} \simeq M_A^2$,
whereas $\widetilde{M}^2_{S,22}$ is bounded from above by eq.~(\ref{tree}). The
element $\widetilde{M}^2_{S,33} \equiv M^2_{SS}$ corresponds to the mass squared
of the CP even singlet state in the decoupling limit $\lambda \to 0$, that can
also be much larger than $M_Z$ if (using eq.~(\ref{softm}))
\beq
M^2_{SS} \simeq \nu (4\nu + A_\kappa) \gg M_Z^2\ .
\label{mss}
\eeq
This leads to the appearance of another potentially large logarithm $\Ls =
\ln(M_{SS}^2/M_Z^2)$.

Now the CP even mass matrix $\widetilde{M}^2_{S}$ can be diagonalized
perturbatively in powers of $M_Z/M_A$ and $M_Z/M_{SS}$. (Recall that these
ratios have to be small whenever large logarithms appear.) During this
perturbative diagonalization we have to keep all terms of ${\cal O}(M_Z^4)$ or
larger in the expression $Tr M_S^4$, that appears as coefficient of potentially
large logarithms in the effective potential~(\ref{cw}). 

A priori this leads to quite lengthy expressions since, for $M_A \gg M_Z$, the
off-diagonal elements $\widetilde{M}^2_{S,13}$ and  $\widetilde{M}^2_{S,23}$ are
of ${\cal O}(M_A M_Z)$. The absence of a negative eigenvalue of
$\widetilde{M}^2_{S}$ (that would signal an instability of the tree level
potential) leads, however, to some simplifications: The subdeterminant in the
$(2,3)$-sector, given by 
$\widetilde{M}^2_{S,22}\widetilde{M}^2_{S,33}-\widetilde{M}^4_{S,23}$, must be
positive. Iff $\widetilde{M}^2_{S,23}\sim {\cal O}(M_A M_Z) \gg M_Z^2$, it
follows from $\widetilde{M}^2_{S,22}\sim {\cal O}(M_Z^2)$ that
$\widetilde{M}^2_{S,33}\gsim M_A^2$ and hence $M_Z/M_{SS}$ is at least as small
as $M_Z/M_A$. On the other hand, recalling the definitions of $\mu = \lambda s$
and $\nu = \kappa s$, one finds that $M_{SS} \gg M_A$ is possible only for
$\lambda \ll \kappa$, in which case the singlet sector decouples and the
off-diagonal elements $\widetilde{M}^2_{S,13}$ and $\widetilde{M}^2_{S,23}$ can
never be large. The resulting contribution from neutral CP even Higgs loops to
the effective potential will be given below.

The situation in the neutral CP odd sector is similarly complex. Again,
it is convenient to perform a first preliminary approximate
diagonalization $\widetilde{M}^2_P = \widehat{\cal O}^T M_P^2\ 
\widehat{\cal O}$ where $\widehat{\cal O}$ differs from ${\cal O}$ in
eq.~(\ref{o}) by $\sb \to -\sb$. 

Let us now review the potentially large matrix elements (with respect 
to $M_Z^2$).
The element $\widetilde{M}^2_{P,11}$ is of ${\cal O}(M_A^2)$ in
the limit (\ref{malim}). The element $\widetilde{M}^2_{P,33}$
corresponds to the mass $M_{PS}$ squared of the CP odd singlet state in
the decoupling limit $\lambda \to 0$. $M_{PS}$ can be much larger than
$M_Z$ if
\beq
M^2_{PS} \simeq -3 \nu A_\kappa \gg M_Z^2
\label{mps}
\eeq
(hence $-3 \nu A_\kappa$ must be positive in this limit). Finally the
element $\widetilde{M}^2_{P,13}$ can also be large. Next we proceed in
two steps: 

First, we semi-diagonalize the $3 \times 3$ matrix 
$\widetilde{M}^2_{P}$ into the above potentially large $2 \times 2$
submatrix and the Goldstone sector, perturbatively in powers of $M_Z$
as in the CP even case. The result is not quite the same as in the CP
even sector, however: In the CP odd case the matrix element
$\widetilde{M}^2_{P,23}$ is never $\gg M_Z^2$. This leads to the
absence of certain terms $\sim \widetilde{M}^2_{P,23}$ as compared to
the CP even case.

Second, we diagonalize exactly the potentially large $2 \times 2$
submatrix consisting of $\widetilde{M}^2_{P,11}$, 
$\widetilde{M}^2_{P,33}$ and $\widetilde{M}^2_{P,13}$. The
corresponding eigenvalues will be denoted by $E_+^2$ and $E_-^2$. (Note
that, for  $M_{PS}^2 \lsim M_A^2$, we have $E_+^2 \sim M_A^2$.)
Potentially large logarithms are now $\ln(E_+^2/M_Z^2)$ and
$\ln(E_+^2/E_-^2)$. The second logarithm is large only for $E_+^2 \gg
E_-^2$, in which case we have $E_+^2 \sim \widetilde{M}^2_{P,11} +
\widetilde{M}^2_{P,33}$,  $E_-^2 \sim
(\widetilde{M}^2_{P,11}\widetilde{M}^2_{P,33} - 
\widetilde{M}^4_{P,13})/E_+^2$ and the rotation angle $\gamma$
satisfies $\sin^2\gamma \sim \widetilde{M}^2_{P,33}/E_+^2$.

We will now present, as an intermediate result, the leading logarithmic
contributions to the effective potential (\ref{cw}) from Higgs boson loops in
terms of the Higgs boson mass matrix elements. (Here $\widetilde{M}_\pm^2$
denotes the charged Higgs mass matrix after the rotation by ${\cal O}_{2\times
2}$, with ${\cal O}_{2\times 2}$  given by the upper left part of ${\cal O}$ in
eq.~(\ref{o}).) The list of all potentially large logarithms is given by
\bea
\La &=& \ln(M_A^2/M_Z^2)\nn \\
\Ls &=& \ln(M_{SS}^2/M_Z^2)\nn \\
\Lmas &=& \ln(\mathrm{max}(M_A^2,M_{SS}^2)/M_Z^2)\nn \\
\Lep &=& \ln(E_+^2/M_Z^2)\nn \\
\Lepm &=& \ln(E_+^2/E_-^2)\ .
\eea

Then we obtain
\bea
\Delta V_{\mathrm{eff}}^{\mathrm{bos}} &=& \frac{1}{64\pi^2}\left(
\left[ 2\widetilde{M}_{\pm,11}^4 + 4\widetilde{M}_{\pm,12}^4 +
\widetilde{M}_{S,11}^4 + 2\widetilde{M}_{S,12}^4 \right] \La 
\right.\nn \\
&&\left. +\left[\widetilde{M}_{S,33}^4 + 2\widetilde{M}_{S,23}^4
+2\widetilde{M}_{S,23}^4\widetilde{M}_{S,22}^2/M_{SS}^2
-\widetilde{M}_{S,23}^8/M_{SS}^4\right]\Ls\right.\nn \\
&&\left. +\left[\widetilde{M}_{P,11}^4 +\widetilde{M}_{P,33}^4 +
2\widetilde{M}_{P,12}^4 + 2\widetilde{M}_{P,13}^4 + 
2\widetilde{M}_{P,23}^4 \right]\Lep\right.\nn \\
&&\left. -\left[E_-^4 +2(\sin\gamma\widetilde{M}_{P,12}^2 -
\cos\gamma\widetilde{M}_{P,23}^2)^2\right]\Lepm
+ 2 \widetilde{M}_{S,13}^4\Lmas
\right)\ .
\eea
(The terms with $M_{SS}$ in the denominator can
still be of ${\cal O}(M_Z^4)$, given the corresponding sizes of the
off-diagonal matrix elements. These terms originate from the
perturbative diagonalization of the CP even mass matrix.)

Finally, the effect of a heavy Higgs doublet on the modification of the running
of the electroweak gauge couplings below $M_A$ can be written in terms of a
contribution to the effective potential that is the same as in the
MSSM~\cite{habhemp}:
\beq
\Delta V_{\mathrm{run}}^{\mathrm{bos}} = \frac{g^4}{192\pi^2}(h_u^2-h_d^2)^2
(-1+2\stq -2\stqq)\ \La\ .
\eeq

It is straightforward, in principle, to compute the derivatives of  $\Delta
V_{\mathrm{eff}}^{\mathrm{bos}}$ and $\Delta V_{\mathrm{run}}^{\mathrm{bos}}$
w.r.t. $h_u$, $h_d$ and $s$, and hence the contributions to the neutral CP even
Higgs mass matrix, given the explicit expressions (\ref{chargh}), (\ref{scalh})
and (\ref{pscalh}) for the Higgs mass matrix elements. The explicit form of
these contributions to all six matrix elements are very lengthy, however.
Subsequently we confine ourselves to the contributions to the mass squared of
the lightest neutral CP even Higgs boson obtained from $\Delta
V_{\mathrm{eff}}^{\mathrm{bos}}$ and $\Delta
V_{\mathrm{run}}^{\mathrm{bos}}$. These simplify considerably, and can
be written as
\bea
&&\Delta^{\mathrm{bos}} M_h^2 = \frac{v^2}{16\pi^2}\left( 
\left[\frac{g^4}{3}\left(\stqq(4+2\sbbq)-\stq(4+8\sbq)
\right.\right.\right.\nn \\
&&\left.\left.\left. -\frac{33}{4}\sbbqq+16\sbbq+\frac{5}{4}\right)
+g^2\lambda^2
\left(2\stq\sbbq+\frac{11}{2}\sbbqq-\frac{15}{2}\sbbq-1\right)
\right.\right.\nn \\
&&\left.\left.
+\lambda^4\left(-\frac{11}{4}\sbbqq+\frac{5}{4}\sbbq+1\right)
\right]\ \La  \right.\nn \\
&&\left. +\left[\lambda^2(\lambda -\kappa\sbb)^2
+\frac{3\lambda^2}{M_{SS}^2} (g^2\cbbq+\lambda^2\sbbq)
(2\mu-\sbb(A_\lambda +2\nu))^2 \right.\right.\nn \\
&&\left.\left. -\frac{\lambda^4}{M_{SS}^4}(2\mu-\sbb(A_\lambda
+2\nu))^4\right]\ \Ls \right.\nn \\
&&\left. +\left[\frac{g^4}{4}(1-\sbbqq) +g^2\lambda^2(\frac{1}{2}\sbbqq
+\frac{1}{2}\sbbq-1) \right.\right.\nn \\
&&\left.\left. +\lambda^4(-\frac{1}{4}\sbbqq-\frac{1}{2}\sbbq+1) 
+\lambda^2(\lambda+\kappa\sbb)^2\right]\ \Lep \right.\nn \\
&&\left. +\left[ \frac{\lambda^2-g^2}{2} \sbbq \cbbq
\frac{M_{PS}^2}{E_+^2}
\right.\right.\nn \\
&&\left.\left. -\left(M_A^2\lambda(\lambda+\kappa\sbb) 
-\lambda^2(A_\lambda -2\nu)^2
-\frac{M_{PS}^2}{2}(g^2\cbbq-\lambda^2(1+\cbbq))\right)^2
/E_+^4
\right.\right. \nn \\
&&\left.\left. +M_A^2 M_{PS}^2 \lambda(\lambda+\kappa\sbb) 
(g^2\cbbq-\lambda^2(1+\cbbq))/E_+^4 \right]\ \Lepm\right)\ .
\label{dmbos}
\eea
$M_A$, $M_{SS}$ and $M_{PS}$ are given in eqs. (\ref{ma}), (\ref{mss})
and (\ref{mps}), respectively. We recall that the terms with $M_{SS}$
or $E_+$ in the denominators are relevant in the limit  $M_{SS},\ E_+
\gg M_Z$ only and, in any case, are valid only if all eigenvalues of
the CP even and odd tree level mass squared matrices are positive. For
$\lambda \to 0$ the result of eq.~(\ref{dmbos}) coincides again with
the one in ref.~\cite{habhemp} obtained in the MSSM.

The contribution (\ref{dmbos}) to  $\Delta M_h^2$ can be translated
back to contributions to the CP even neutral Higgs mass matrix elements
according to the rules
\bea
\Delta^{\mathrm{bos}} M_{uu}^2 &=& \sbq \Delta^{\mathrm{bos}} M_h^2\nn \\
\Delta^{\mathrm{bos}} M_{dd}^2 &=& \cbq \Delta^{\mathrm{bos}} M_h^2\nn \\
\Delta^{\mathrm{bos}} M_{ud}^2 &=& \sb \cb \Delta^{\mathrm{bos}} M_h^2
\eea
which gives the correct results in the limit where both $M_A$ and
$M_{SS}$ are much larger than $M_Z$, and the corrections to the masses
of the heavy Higgs states  are negligible. (The resulting expressions
for $\Delta^{\mathrm{bos}} M_{uu}^2\ , \Delta^{\mathrm{bos}} M_{dd}^2\
,\Delta^{\mathrm{bos}} M_{ud}^2$ differ from the ones in
ref.~\cite{habhemp}, i.e. our method gives different corrections to the
mass of the heavy Higgs states.)

For completeness we give also the contribution of sfermions (squarks and
sleptons) $\sim g^4$ to the mass of the lightest CP even neutral Higgs boson,
which are again the same as in the MSSM. As in ref.~\cite{habhemp} we assume a
common mass $M_{\mathrm{susy}}$ for all squarks and sleptons, with the exception
of the squarks of the third generation whose masses are denoted by $m_{Q}$,
$m_{T}$ and $m_{B}$. Defining
\bea
&&\Delta^{\mathrm{sferm}} = \frac{v^2 g^4}{12\pi^2} \left\{
\left(\frac{3}{2}-3\stq+\frac{5}{3}\stqq\right)\ln(m_Q^2/M_Z^2) \right . \nn \\
&& \left. +\frac{4}{3}\stqq \ln(m_T^2/M_Z^2)  + \frac{1}{3}\stqq
\ln(m_B^2/M_Z^2)  \right. \nn \\
&& \left. +\left(\frac{9}{2} - 9\stq + \frac{38}{3}\stqq\right)
\ln(M_{\mathrm{susy}}^2/M_Z^2) \right\} 
\label{dmsfermuniv}
\eea
one finds
\bea
\Delta^{\mathrm{sferm}} M_{uu}^2 &=& \sbq \Delta^{\mathrm{sferm}} \nn \\
\Delta^{\mathrm{sferm}} M_{dd}^2 &=& \cbq \Delta^{\mathrm{sferm}} \nn \\
\Delta^{\mathrm{sferm}} M_{ud}^2 &=& -\sb \cb \Delta^{\mathrm{sferm}} 
\label{dmsferm}
\eea
in agreement with ref.~\cite{habhemp}.

\section{Impact on the Mass of the Lightest Neutral Higgs Boson} 
\label{sec:imp}

In the present paper we have computed all leading logarithmic radiative
corrections to the lightest Higgs mass in the NMSSM, to fourth order in the
electroweak gauge couplings, $\lambda$ and $\kappa$ (including all mixed
combinations). The essential results are the radiative corrections
(\ref{dmferm}) and (\ref{dmbos}) to the mass squared of the lightest CP even
neutral Higgs boson induced by chargino, neutralino and Higgs boson loops. The
numerical impact on its mass $\Delta M_h$ can most easily be studied using
\beq
\Delta M_h \simeq \frac{1}{2 M_h} \Delta M_h^2\ .
\label{dm}
\eeq

Subsequently we want to plot the numerical effect of the radiative
corrections considered here as a function of $\tan\!\beta$ or
$\lambda$. However, since $M_h$ depends on $\tan\!\beta$ and $\lambda$
already at tree level, the straightforeward use of eq. (\ref{dm}) to
compute $\Delta M_h$ would not allow to disentangle the effect due to
the radiative corrections from the effect due to the tree level
variation of $M_h$. In addition, radiative corrections from the
top/stop sector on $M_h$ in the denominator of eq. (\ref{dm}), which we
do not consider in this paper, are also
very important in the NMSSM. Therefore, for the purpose to focus on
the effect due to the radiative corrections in eq. (\ref{dm}), we fix
the numerical value of $M_h$ in the denominator to $M_h=120$~GeV, just
above the lower limit from LEP. The numerical values for $\Delta M_h$
obtained below can easily be rescaled to any other value of $M_h$ using
the relation above. (Notably they will decrease for larger values of
$M_h$.)

\begin{figure}[htb]
\begin{center}
\vspace*{6mm}
\epsfig{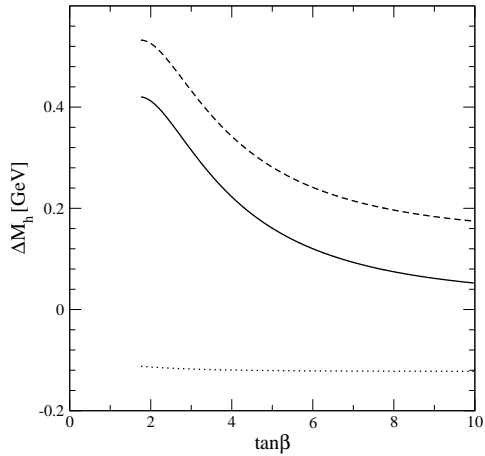}
\vspace*{-6mm}
\end{center}
\caption{$\Delta^{\mathrm{ferm}} M_h$ (dotted line),
$\Delta^{\mathrm{bos}} M_h$ (dashed line) and the sum (full line) as a
function of $\tan\!\beta$ for $\lambda = 0.1$, $\kappa = 0.5$, $\mu =
M_2 = 100$~GeV, $A_\lambda = 1$~TeV and $A_\kappa = -100$~GeV.}
\end{figure}

\begin{figure}[htbp]
\begin{center}
\vspace*{6mm}
\epsfig{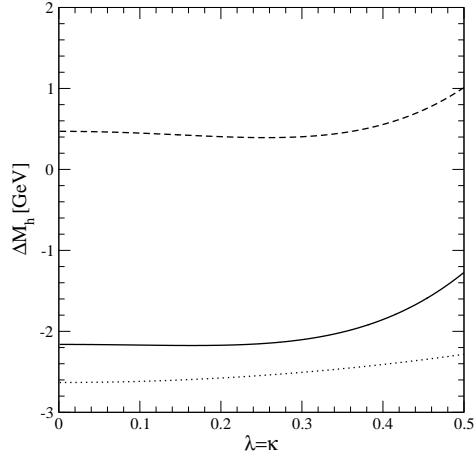}
\vspace*{-6mm}
\end{center}
\caption{$\Delta^{\mathrm{ferm}} M_h$ (dotted line),
$\Delta^{\mathrm{bos}} M_h$ (dashed line) and the sum (full line) as a
function of $\lambda = \kappa$ for  $\tan\!\beta =1.5$, $\mu =
500$~GeV, $M_2 = 1$~TeV, $A_\lambda =  -250$~GeV, and $A_\kappa =
-1$~TeV }
\end{figure}

First we are interested in maximizing the upper bound on the
lightest Higgs mass. Whereas the fermionic contributions (\ref{dmferm}) 
to $M_h^2$ are generally negative (and small for chargino and neutralino
masses $\sim M_Z$), the bosonic contribution (\ref{dmbos}) can be
marginally positive. The following choice of parameters (for soft terms
bounded by 1 TeV, and $\lambda$ and $\kappa$ bounded by the absence of
a Landau pole below $M_\mathrm{GUT}$) renders the sum of the fermionic
and bosonic contributions as positive as possible:
$\lambda = 0.1$, $\kappa = 0.5$, $\mu = M_2 = 100$~GeV, $A_\lambda =
1$~TeV and $A_\kappa = -100$~GeV. In Fig.~1 we plot
$\Delta^{\mathrm{ferm}} M_h$, $\Delta^{\mathrm{bos}} M_h$ and the sum
as a function of $\tan\!\beta$. Values of $\tan\!\beta \lsim 1.5$ are
omitted since they would give raise to a Landau Pole for the top quark
Yukawa coupling. We see that the maximal increase in $M_h$ is just $\sim
0.5$~GeV.

\begin{figure}[htbp]
\begin{center}
\vspace*{6mm}
\epsfig{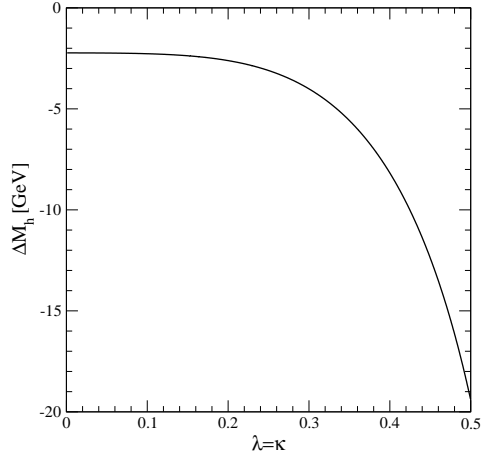}
\vspace*{-6mm}
\end{center}
\caption{Total contribution to  $\Delta M_h$  as a function of $\lambda
= \kappa$ for $\tan\!\beta =2$, $\mu = M_2 = 1$~TeV, $A_\lambda = 
-800$~GeV, and $A_\kappa = -50$~GeV }
\end{figure}

In order to see the impact of the NMSSM specific contributions 
$\sim~\lambda$ and $\sim~\kappa$, we plot in Fig.~2 $\Delta M_h$ as a
function of $\lambda$ (keeping $\kappa= \lambda$) for $\tan\!\beta =$
1.5. The other parameters are such that the positive NMSSM specific
effect is maximized: $\mu = 500$~GeV, $M_2 = 1$~TeV, $A_\lambda = 
-250$~GeV, and $A_\kappa = -1$~TeV. Values of $\lambda$ larger than 0.5
would correspond to a Landau Pole below the GUT scale. Here one sees
that the NMSSM specific contributions to $M_h$ (the difference between
$\lambda = \kappa = 0.5$ and $\lambda = \kappa = 0$) are marginally
positive, up to +1~GeV within the allowed range for the Yukawa
couplings. 

However, the NMSSM specific contributions to the lightest Higgs mass
can also be negative, and some of them can be quite large in absolute
value. These contributions originate from the CP odd sector and are of
the form $-(v^2/16\pi^2)\lambda^4 (A_\lambda-2\nu)^4/E_+^4\ \Lepm$
(obtained by expanding the square of the next-to-last line in eq.
(\ref{dmbos})). If $|A_\lambda-2\nu| \gg E_+$ (keeping $E_+ > M_Z$ and a
positive determinant for the CP odd mass matrix), this
contribution can lower the lightest Higgs mass by up to $\sim -20$~GeV.
The corresponding phenomenon is shown in Fig.~3, which gives the total
contribution to  $\Delta M_h$  as a function of $\lambda = \kappa$ for
$\tan\!\beta =2$, $\mu = M_2 = 1$~TeV, $A_\lambda =  -800$~GeV, and
$A_\kappa = -50$~GeV.

To conclude, the radiative corrections to the lightest Higgs mass in
the NMSSM considered here are typically negative (as large as
$-20$~GeV, if $A_\lambda$ and $\nu$ differ in sign), but can increase its
upper bound only by up to +1~GeV.

\end{document}